\documentclass[manuscript,screen]{acmart}

\AtBeginDocument{%
  \providecommand\BibTeX{{%
    \normalfont B\kern-0.5em{\scshape i\kern-0.25em b}\kern-0.8em\TeX}}}

\copyrightyear{2021}
\acmYear{2021}

\acmConference[SimuRec '21]{Workshop on Simulation Methods for Recommender Systems at ACM RecSys '21}{October 1st, 2021}{Amsterdam, Netherlands}
\acmBooktitle{Workshop on Simulation Methods for Recommender Systems at ACM RecSys '21 (SimuRec '21), October 1st, 2021, Amsterdam, Netherlands}

\makeatletter
\renewcommand\@formatdoi[1]{\ignorespaces}
\makeatother

\usepackage[inline]{enumitem}

\begin{document}

\title[Simulation for Reward-Optimizing Recommender Systems]{Offline Evaluation of Reward-Optimizing Recommender Systems: The Case of Simulation}
\subtitle{A Position Paper}

\author{Imad Aouali}
\author{Amine Benhalloum}
\author{Martin Bompaire}
\author{Benjamin Heymann}
\author{Olivier Jeunen\textsuperscript{1}}
\author{David Rohde}
\author{Otmane Sakhi}
\author{Flavian Vasile}
\authornote{Authors listed in alphabetical order.}

\affiliation{%
  \institution{Criteo}
  \city{Paris}
  \country{France}
}

\email{}
\affiliation{%
  \institution{\textsuperscript{1}Adrem Data Lab, University of Antwerp}
  \city{Antwerp}
  \country{Belgium}
}


\renewcommand{\authors}{Imad Aouali, Amine Benhalloum, Martin Bompaire, Benjamin Heymann, Olivier Jeunen, David Rohde, Otmane Sakhi and Flavian Vasile}
\renewcommand{\shortauthors}{}

\begin{abstract}
Both in academic and industry-based research, online evaluation methods are seen as the golden standard for interactive applications like recommendation systems.
Naturally, the reason for this is that we can directly measure utility metrics that rely on interventions, being the recommendations that are being shown to users.
Nevertheless, online evaluation methods are costly for a number of reasons, and a clear need remains for reliable offline evaluation procedures.
In industry, offline metrics are often used as a first-line evaluation to generate promising candidate models to evaluate online.
In academic work, limited access to online systems makes offline metrics the de facto approach to validating novel methods.
Two classes of offline metrics exist: proxy-based methods, and counterfactual methods.
The first class is often poorly correlated with the online metrics we care about, and the latter class only provides theoretical guarantees under assumptions that cannot be fulfilled in real-world environments.
Here, we make the case that simulation-based comparisons provide ways forward beyond offline metrics, and argue that they are a preferable means of evaluation.
\end{abstract}

\maketitle

\section{Introduction}
Current best practice in recommender systems evaluation, both in academic and industrial settings, is based on using real datasets to compute \emph{offline} metrics that are proxies to \emph{online} performance.
In academic work, these metrics are used to validate the effectiveness of newly proposed approaches in the literature.
In industry, if a sufficient number of offline metrics are promising, the method is further tested in online experiments such as A/B-tests.
In this paper we argue: \emph{this is a sub-optimal methodology}.
The proxies that we use are too poorly correlated with online performance to give a reasonable measurement of how likely that a candidate model succeeds in an online experiment. 
We propose an alternative methodology where instead of using offline proxy metrics, we simulate user behaviour and evaluate whether the recommender system is able to generate \emph{good} recommendations on simulated timelines.
The advantage of simulations is that we can measure estimates of actual \emph{online} reward, instead of needing to resort to offline proxies with their widely reported flaws.

The dilemma at the heart of this issue is:
\begin{enumerate*}
\item should we use proxies that are quite poorly correlated with actual performance and consequently suffer from Goodhart's law: ``Any observed statistical regularity will tend to collapse once pressure is placed upon it for control purposes'', or
\item should we use simulated timelines that allow actual calculation of reward but are by virtue of being simulation not representative of a real system (although possibly informed by it).
\end{enumerate*}
In this paper we make the case in favor of simulation.

In Section 2, we discuss the disadvantages of proxy metrics such as Recall@K and click-rank and how they can be poor proxies to actual performance.
In Section 3, we discuss why reward-optimizing recommendation using inverse propensity scoring (IPS) is not viable in practice.
In Section 4, we argue that the ability of a recommender system to optimize the reward in a simulation environment such as RecoGym and RecSim \citep{rohde2018recogym,ie2019recsim} is a more compelling case that it will perform well in production. Section 5 concludes our argument.

\section{Proxy methods are not trustworthy indicators of performance}
Over the past decade or so, the focus of recommendation research has moved on from the classical \emph{rating prediction} task to that of \emph{next-item prediction}~\cite{Steck2013}.
Here, the goal of the recommender is to complement sequences of observed organic user-item interactions with other items that the user may find relevant.
Following classical paradigms in evaluation of supervised learning methods, certain items are obfuscated to make up the ground truth of the test set.

Information Retrieval-inspired metrics such as Recall, Normalised Discounted Cumulative Gain, Mean Average Precision and others are then computed on top-$K$ lists of generated recommendations, and used to evaluate which systems are better at ranking the obfuscated items higher than others.
Such metrics result in having an idealized list of recommendations produced with a heuristic \cite{Benhalloum21}.
Time and again, research has shown that this type of evaluation procedure does not yield results that are sufficiently correlated with the results from randomised control trials (which we see as the golden standard)~\cite{Garcin2014,Rossetti2016,Jeunen2019_DS}.
While these methods are powerful and have fostered impressive research progress over the years, they remain a compromise.
This mismatch between offline and online evaluation results in turn leads to a rift between academia and industry, which increasingly rely on these two respective alternatives.
Several reasons can be cited for this disparity, such as temporal constraints in the data not being adhered to~\cite{Jeunen2018}, data leakage~\cite{ji2021critical} or a lack of clearly defined best practices~\cite{Dacrema2019}.
Additionally, we argue that these offline metrics do not measure the same signal as the online metrics that we care about: they are but proxies~\cite{Jeunen2019REVEAL_EVAL}.

Indeed, this setting still does not accurately reflect the recommendation use-case practitioners face in industry.
In practice, we care about some notion of \emph{reward} that we wish to maximise.
The probability that a recommendation yields some reward is a function that maps the cross of a \emph{user timeline} and a list (or slate) of recommendations to a reward.
Optimising this directly is usually not viable in real-world systems, because it isn't possible to accurately measure the reward function. Not only do proxy methods not correlate with performance they do not have the same support. 

If our online evaluation metrics (such as click-through-rate (CTR)) rely on interventions (shown recommendations), our offline metrics should equally make use of interventional data.
Moving away from proxies -- we can then adopt \emph{offline} counterfactual estimators for these \emph{online} metrics, and make meaningful assertions about the projected performance of a system.
Nevertheless, counterfactual estimation methods are no silver bullet, and have several problems that are difficult to fully mitigate, as we discuss in the following Section.

\section{Inverse Propensity Scoring Methods are not viable in production}


Another approach that has been popular academically but has had relatively little impact in production systems is to use the inverse propensity scoring (IPS) \cite{bottou2013counterfactual} to design unbiased estimators of the reward, that we denote by $\hat{V}_n(\cdot)$. Precisely, we have access to logged data $\mathcal{D}_n = \{ x_i, a_i, r_i\,, \ i=1, \ldots, n\}$ collected by a
logging policy $\pi_0$ where samples $(x_i, a_i, r_i)$ are drawn independently as $(x, a, r) \sim \nu(x) \pi_0(a \mid x) p(r \mid x, a)$. Here $\nu(\cdot)$ is a distribution over the contexts and $p(\cdot \mid x, a)$ is the distribution of the reward given $x$ and $a$. IPS-based methods remove the preference bias of the logging policy $\pi_0$ in logged data $\mathcal{D}_n$ by re-weighting samples using the discrepancy between the target policy $\pi$ and the logging policy $\pi_0$. A typical estimator $\hat{V}_n(\cdot)$ has the form
\[
\hat{V}_n(\pi) = \frac{1}{n} \sum_{i=1}^n r_i \frac{\pi(a_i|x_i)}{\pi_0(a_i|x_i)}.
\]
The target policy is often parametrized as $\pi_\beta$, and can be optimized to place high mass on historical actions that resulted in positive reward. Now, consider the following ad-placement example where the reward is a click indicator. Here, we assume that we have $10^6 + 1$ items to recommend and the context $x$ is discrete and has $10^3$ equally probable states. We also assume that the logging policy is epsilon-greedy with $\epsilon=0.01$, and that we have a large dataset with $n=10^9$. Finally, assume that the best action has a CTR of 2\% but a poor action has a CTR of 1\%.

For a particular context $x$, we have observed $10^6$ impressions. Of those, $10^4$ will be used for exploration (the rest exploit the best arm as estimated by the logging policy). The $10^4$ exploration steps will be distributed over the $10^6$ other actions. Even if all of these actions have a CTR of $1$\%, we will then see around $100$ clicks on actions that we have usually only tried a single time.
The IPS estimator will estimate an illegal CTR of $100=\frac{1}{0.01} =10\,000\%$, i.e. a CTR much greater than $1$.
IPS extensions like weight capping~\cite{Gilotte_2018} might be implemented in order to make the estimate legal, but the capping parameter will dictate whether estimates that observe one out of one click are deemed to be better than the existing production system.

We've made the simplifying assumption that a single recommendation is delivered at a time, which is often unrealistic in real-world systems.
When dealing with lists (or slates) of items $a_1, \ldots, a_K$, an action space of size of $10^6 + 1$ is extremely small. IPS-based methods do not handle these settings well, as they are based upon counts of exact matches of contexts and actions.
In order to alleviate this sort of problem, \citeauthor{chen2019top} make the assumption that only one recommendation has ongoing impact \cite{chen2019top}.
This is functionally equivalent to administering multiple drugs to a patient, taking a measurement, and on the basis of this measurement ignoring all but one of the administered drugs in the measurement of any effects. 
In short. it violates the protocols of randomized control trials in a very egregious way.
This isn't to say this assumption may not be useful, but it is simply a departure from reward-optimizing recommendation. 
For other similar approaches in a slate setting, see \cite{li2018offline, swaminathan2017offpolicy, McInerney2020}.
\citeauthor{Gilotte_2018} provide an excellent survey of variance issues with IPS methods in the context of recommendation \cite{Gilotte_2018}.

It is important to note that it is \emph{impossible} to borrow strength or reduce estimation variance by restricting the parametric form of $\pi(a_1,...,a_K|x)$.
Borrowing strength amounts to assuming that $p(r|a_1,...,a_K,x)$ is correlated with $p(r|a_1',...,a_K',x')$ if $a_1,...,a_K$ is close to $a_1',...,a_K'$ and $x$ is close to $x'$.
Implementing these assumptions is possible using slate models \cite{Aouali21}, and the fundamental distances of bandit recommendation \cite{Sakhi2020}.
In contrast, restricting $\pi(a_1,...,a_K|x)$ does nothing to reduce the variance of the reward of a given distribution within $\pi$, as only exact coincidence between $a_1,...,a_K$ and $x$ can be used.



\section{Simulation offers a partial path to reward optimizing recommendation}
We have argued so far that it isn't feasible to use proxy offline metrics to measure performance of real-world systems.
Our experience is that, although these are widely used for \emph{candidate} generation of promising recommendation models, they are not fully trusted.
In academic research, these offline metrics are often the only available tool for experimental validation, but several recent works have cast doubt on their utility, and there is a growing consensus that they should not be taken at face value.
In many industrial settings, the output of these models is manually validated by an editorial team before they are progressed to online tests. 

In practice, a chosen recommender system candidate model may not outperform on the offline metrics but, is put forward because it appears to produce sensible output.
In other words, Goodhart's law is assumed to be operating.

We further argued that IPS style counterfactual estimators that -- in contrast to proxies -- should be correlated to reward, are of little use in practical settings.
They have been a source of intense academic study, but have had little broad practical impact.

We now introduce the possibility of simulation-based evaluation  as an alternative. At the outset, we need to state that simulation has an obvious downside -- it is likely to differ from the real world, but we need to also keep in mind the downsides of proxy based approaches and counterfactual based approaches.

There are two broad ways we could use simulation to validate recommender systems:
\begin{enumerate*}
    \item Seed the simulator many times and compute actual performance metrics averaging over the many simulation runs.
    This sort of simulation has the form of a basic sanity check.
    Was the system able to exploit a recommendation signal, did the learning method manage to infer a sensible model of the data-generating process?
    It can be viewed as a test in software engineering, or as measuring an estimator's performance in statistics. 
    If good performance is obtained, the recommender system is trained on the actual logs after validation in the simulation environment.
    \item Use a simulator that is informed by past recommendations and a modelling approach.
    The simulator will use modelling to answer as accurately as possible the expected reward of new recommendations.
    In this case, when a good recommender system is discovered, it will be deployed directly.
    At its most idealistic, this approach is compatible with Bayesian decision theory.
\end{enumerate*}
Both approaches may have value, likely as stated (2) is too idealistic but going beyond (1) will likely have further value.

Adopting either of these approaches consistently would require a paradigm shift in the recommender systems community, and it is speculative to say at this stage if it will work.
Likely a transition from metrics to simulation could only be achieved gradually, perhaps after studies comparing simulation and offline metrics as candidate generators where A/B-tests are the final arbiter.

An interesting counterargument to the use of simulation methods, is that a good simulator requires an understanding of user behaviour that is just as complex as the recommender system itself.
With the first approach -- we depart from the strict view that the simulator must be an extremely accurate representation of a real-world system.
In contrast, we view the simulator as a data generating process -- and we wish to measure which learning method can model the data generating process well enough such that it can accumulate a higher reward in future interactions.
Even though the data generating process is not entirely the same as the one in the natural world, we can say with some confidence that the performance of learning methods might transfer from one to the other.

Using the second approach, we must answer: \emph{How do you know the simulator is good enough?}
This is different to the usual view of performance that is adopted in machine learning and recommender systems.
We are not measuring performance, but rather fidelity.
Ideas such as posterior predictive tests may help \cite{gelman1995bayesian}.

In academia, the case for simulation is both more straight forward and more difficult.
Simulation studies now have a track record of showing the viability of recommendation algorithms in ways that go beyond what is possible with offline data sets \cite{Mykhaylov2019CausalML,Jeunen2019REVEAL,Sakhi2020,JeunenKDD2020,Jeunen2020REVEAL,Jeunen2021,Jeunen2021B,Bendada2020}.
On the other hand, the inertia created by the \emph{MovieLens} tradition in recommender systems will be displaced only very slowly, and even though offline metrics are flawed, they have clearly shown their value in the past decades of research progress.

\section{Conclusion: the case for simulation}

A recommender system usually consists of a relatively simple piece of engineering - a personalized ranker of items.  This simple piece of engineering interacts with a complex world of users both interacting independently of the recommender system and at other times receiving sequences of slates of recommendations.
Furthermore the owner of the recommender system may have quite complex goals: for example, to encourage long term engagement of users or to drive sales.

The current state of the art approach is to dodge this complexity and formulate a machine learning problem as a distance between items and users (usually summarized as a sequence of items), and sometimes as simply a distance between items and items.
Both of these approaches massively simplify what the system does and raise technical questions that are arguably impossible to answer.
When we formulate the problem in these terms, we limit how we see the world by adopting the limitations of implementable decision rules.

While producing accurate simulation is obviously fraught - simulation remains one of the few ways of attempting to do true reward-optimizing recommendation.
Whether this approach can really replace the massive reductionism of item-user distance or item-item distance is unclear at this stage, but it must be noted that simulation is one of the very few ways we have available to tackle the central open question of recommender systems: \emph{How can we build true reward-optimizing recommender systems?}

\bibliographystyle{ACM-Reference-Format}
\bibliography{sample-base}

\end{document}